\documentclass{amsproc}

\usepackage{amsmath}
\usepackage{amsfonts}
\usepackage{amssymb}
\usepackage{amscd}
\usepackage{amsthm}
\usepackage{latexsym}

\usepackage{exscale}
\usepackage{amsbsy}

 \def\X{\mathcal X} \def\C{\mathbb{C}} 
\def\N{\mathbb{N}} \def\R{\mathbb{R}} \def\T{\mathbb{T}} 
\def\A{{\mathcal A}} \def\B{{\mathcal B}}  
  \def\H{\mathcal H} 
 
\def\S{\mathcal S}  \def\U{\mathcal U}
\def\CCC{{\mathfrak C}} \def\FF{{\mathfrak F}}

\def\Rep{\mathfrak{Rep}}  \def\de{\mathrm{d}}  
   
\def\im{\mathop{\mathsf{Im}}\nolimits} 
\def\I{{\rm 1\kern-.26em I}}
\def\SA{S_{\!\A}} \def\FA{F_{\!\A}}

\def\oB{\omega^{\hbox{\it \tiny B}}}

\def\var{\mathfrak z} 

\def\CBA{{\mathfrak C}^{B}_{\!\A}} \def\BBA{{\mathfrak B}^{B}_{\!\!\A}}
\def\a{\mathfrak a}
\def\b{\mathfrak b}
\def\T{\X^*,\A^\infty}
\def\Q{\mathbf Q}

 \def\ad{\mathfrak{ad}}

\def\1{\mathfrak{1}}
\def\0{\mathfrak{0}}
\def\G{\mathfrak{G}}

\newtheorem{theorem}{Theorem}[section]
\newtheorem{lemma}[theorem]{Lemma}

\theoremstyle{definition}
\newtheorem{definition}[theorem]{Definition}

\newtheorem{corollary}[theorem]{Corollary}
\newtheorem{proposition}[theorem]{Proposition}
\newtheorem{hypothesis}[theorem]{Hypothesis}

\theoremstyle{remark}
\newtheorem{remark}[theorem]{Remark}

\numberwithin{equation}{section}

\begin{document}

\title{The magnetic formalism; new results}

\author{Viorel Iftimie}
\address{Institute of Mathematics Simion Stoilow of the Romanian Academy, P.O.  Box
1-764, Bucharest, RO-70700, Romania}
\email{Viorel.Iftimie@imar.ro}
\thanks{Viorel Iftimie is partially supported from Contract No.2 CEx06-11-18/2006}
\author{Marius M\u antoiu}
\address{Departamento de Matem\'aticas, Universidad de Chile, Las Palmeras 3425, Casilla 653,
Santiago, Chile }

\email{Marius.Mantoiu@imar.ro}
\thanks{Marius M\u antoiu was supported by the Fondecyt Grant No.1085162 and by the N\'ucleo Cientifico ICM P07-027-F "Mathematical
 Theory of Quantum and Classical Systems".}

\author{Radu Purice}
\address{Institute of Mathematics Simion Stoilow of the Romanian Academy, P.O.  Box
1-764, Bucharest, RO-70700, Romania} \email{Radu.Purice@imar.ro}
\thanks{Radu Purice is partially supported from Contract No.2 CEx06-11-18/2006 and by Laboratoire Europ\'een Associ\'e CNRS Math-Mode.}

\date{January 1, 1994 and, in revised form, June 22, 1994.}



\begin{abstract}
We review recent results on the magnetic pseudodifferential
calculus both in symbolic and in $C^*$-algebraic form. We also
indicate some applications to spectral analysis of
pseudodifferential operators with variable magnetic fields.

\end{abstract}

\maketitle

\section{Introduction}

It is commonly accepted that the Weyl form of the pseudodifferential calculus defines a convenient quantization of a physical system
composed of a non-relativistic particle without spin moving in $\R^n$, when no magnetic field is present. Beyond foundational
matters, the formalism is also very useful for various results and computations.

The problem of defining quantum observables in the presence of a non-homogeneous magnetic field is a non-trivial one.
A naive use of the Weyl calculus fails, missing gauge covariance. It also ignores the fact that a magnetic field changes the geometry
of the phase space in a way that really requires a new pseudodifferential calculus.

A correct solution was offered quite recently
(with various degrees of generality and rigor) in works as \cite{Mu,KO1,KO2,MP1,MPR1,MP2,IMP1}. The first stage of the theory
has been reviewed in \cite{MP3}. We give here a brief presentation of the subsequent development of the subject.

 We work in the $n$-dimensional space $\mathcal{X}:=\mathbb{R}^n$ and in the associated phase space
 $\Xi:=\mathcal{X}\times\mathcal{X}^*\equiv\R^{2n}$, endowed with the canonical symplectic form
$$
\sigma(X,Y)\equiv\sigma((x,\xi),(y,\eta)):=y\cdot\xi-x\cdot\eta.
$$
The magnetic field is described by a closed 2-form $B$ on $\mathcal{X}$, to which we associate a perturbation
of the canonical symplectic form on $\Xi$:
$$
\sigma^B_{(z,\zeta)}((x,\xi),(y,\eta)):=\sigma((x,\xi),(y,\eta))+B(z)(x,y),\ \ (x,\xi),(y,\eta),(z,\zeta)\in\Xi.
$$
The importance of this symplectic form in the classical theory of systems in magnetic fields is outlined in \cite{MR,DR}.
It also defines a Poisson structure on the smooth classical observables which serves as a semiclassical limit for the subsequent
quantum formalism, as shown in \cite{MP2}.

To $B$ we may associate in a highly non-unique way vector potentials, i.e. 1-forms $A$ such that $B=dA$.
 Under gauge transformations $A\mapsto A^\prime=A+d\phi$ (so that $B=dA=dA^\prime$) the quantization should behave covariantly.

Suppose chosen a gauge $A$ for the magnetic field $B$.
We have to define a functional calculus for the family of non-commuting operators
$$Q_1,\ldots,Q_n;\Pi^A_1=D_1-iA_1,\ldots,\Pi^A_n=D_n-iA_n$$
representing the canonical variables ($Q_j$ is the operator of
multiplication by $x_j$ and $D_j=-i\partial_j$).
 We shall use the unitary operators associated to the above $2n$ self-adjoint operators and define
 {\it the magnetic Weyl system}
$$
W^A(x,\xi):=\exp \{-i\sigma[(x,\xi),(Q,\Pi^A)]\},\ \ \ \ \ (x,\xi)\in\Xi.
$$
For functions $f:\Xi\rightarrow\mathbb{C}$ we define the associated magnetic Weyl operator
$$
\mathfrak{Op}^A(f):=\int_\Xi dX\hat{f}(X)W^A(X)
$$
given by the following formula (admitting various interpretations, depending on the properties of
$f$ and $u:\X\rightarrow\mathbb C$):
\begin{equation}\label{op}
\left[\mathfrak{Op}^A(f)u\right](x):=(2\pi)^{-n}\int_\X\int_{\X^*}dy\,d\xi\,e^{i(x-y)\cdot\xi}e^{-i\int_x^y A}
f\left(\frac{x+y}{2},\xi\right)u(y).
\end{equation}
The operators associated to any two gauge-equivalent vector potentials are unitarily equivalent:
$$
A^\prime=A+d\phi\quad\Rightarrow\quad\mathfrak{Op}^{A^\prime}(f)=e^{i\phi(Q)}\mathfrak{Op}^A(f)e^{-i\phi(Q)}.
$$
This is due to our correct choice of the phase factor $\exp (-i\int^y_x A)$, containing the circulation of the vector potential.

In Chapter 2, following mainly \cite{IMP1}, we review properties of the quantization $f\mapsto\mathfrak {Op}^A (f)$, especially for $f$
belonging to some of the H\"ormander classes of symbols $S^m_{\rho,\delta}(\Xi)$. We present boundedness results and describe the
magnetic Sobolev spaces. An important role is played by a well-behaved composition law $\sharp^B$ defined on symbols, only depending on
the magnetic field.

Chapter 3 is devoted to the extension to our magnetic pseudodifferential calculus of the commutator criteria of Beals and Bony. As explained
 in \cite{IMP2}, they have important consequences on the behavior of various classes of magnetic pseudodifferential operators
 under inversion and functional calculus.

In Chapter 4 we present (cf. \cite{MPR1} and \cite{LMR}) the operator algebra version of the magnetic formalism. We recast the basic
information in a twisted
$C^*$-dynamical system, to which one canonically assigns twisted crossed product algebras. An important property of these is to contain the
resolvent families of unbounded magnetic pseudodifferential operators with anisotropic coefficients. This is called affiliation and has
implications in spectral analysis.

The anisotropy is encoded in an abelian $C^*$-algebra whose Gelfand spectrum is a compact dynamical system. The quasi-orbit structure of
this dynamical systems will be shown in 5.2 (cf. \cite{MPR2} and \cite{LMR}) to contain relevant information on the essential
spectrum of affiliated operators.

Another application is a limiting absorbtion principle and the corresponding spectral information for rather general classes of operators
with decaying magnetic fields. This involves Mourre's commutator method and is presented in 5.1 following \cite{IMP1}.

\section{The magnetic pseudodifferential calculus}

\subsection{The composition law}
The functional calculus $f\mapsto\mathfrak{Op}^A(f)$ induces a {\it magnetic composition} on the space of test functions $\mathcal{S}(\Xi)$,
by requiring
$$
\mathfrak{Op}^A(f\sharp^B g)\,:=\,\mathfrak{Op}^A(f)\,\cdot\,\mathfrak{Op}^A(g).
$$
Explicitly we have
$$
(f\sharp^B g)(X):=\pi^{-2n}\int_\Xi dY\int_\Xi dZ\,e^{-i\int_{\mathcal{T}_X(Y,Z)}\sigma^B}\,f(X-Y)\,g(X-Z),
$$
involving the flux of the 2-form $\sigma^B$ through $\mathcal{T}_X(Y,Z)$, the triangle in $\Xi$ having vertices
$$
X-Y-Z,\quad X+Y-Z,\quad X-Y+Z.
$$
Setting
$$
\mathfrak{M}^B(\Xi):=\left\{f\in\mathcal{S}^\prime(\Xi)\mid f\,\sharp^B\phi\in\mathcal{S}(\Xi),\ \phi \ \sharp^B
f\in\mathcal{S}(\Xi),\ \forall\phi\in\mathcal{S}(\Xi)\right\},
$$
we get a $^*$-algebra for the composition law $\sharp^B$ (extend by duality techniques)
and the usual complex conjugation as involution. $\mathfrak{M}^B(\Xi)$ is very large; among others, it was shown in \cite{MP1} that
the space of indefinitely differentiable functions with uniform polynomial growth is contained in $\mathfrak{M}^B(\Xi)$.

Let us denote by $\mathbb B(\mathcal R)$ the family of all continuous, linear operators in the topological vector space $\mathcal R$.
The main property of $\mathfrak{M}^B (\Xi)$ is the fact that $$\mathfrak {Op}^A:\,\mathfrak{M}^B(\Xi)\to\mathbb B[\mathcal S(X)],\ \ \ \
\mathfrak {Op}^A:\,\mathfrak{M}^B(\Xi)\to\mathbb B[\mathcal S'(X)] $$
are one-to-one $*$-representations.

By gauge covariance, the $^*$-algebra
$$
\mathfrak{A}^B(\Xi):=\left\{f\in\mathcal{S}^\prime(\Xi)\mid\mathfrak{Op}^A(f)\in\mathbb{B}[L^2(\mathcal{X})]\right\}
$$
 does not depend on the choice of $A$, but only on the magnetic field $B$.
On $\mathfrak{A}^B(\Xi)$ we define the map
$$
\|f\|_B:=\|\mathfrak{Op}^A(f)\|_{\mathbb{B}[L^2(\mathcal{X})]},
$$
that is in fact a C$^*$-norm on $\mathfrak{A}^B(\Xi)$ only depending on $B$.
Thus $\mathfrak{A}^B(\Xi)$ is a C$^*$-algebra isomorphic to $\mathbb{B}[L^2(\mathcal{X})]$.

\subsection{Magnetic composition of symbols}

For $m\in\mathbb{R}$, $0\leq\delta\leq\rho\leq 1$ and $f\in C^\infty(\Xi)$, we introduce the family of seminorms
$$
|f|_{(a,\alpha)}^{(m;\rho,\delta)}:=\underset{(x,\xi)\in\Xi}{\sup}<\xi>^{-m+\rho|\alpha|-\delta|a|}\left|(\partial_x^a\partial_\xi^
\alpha f)(x,\xi)\right|,\;
$$
and define the H\"ormander symbol classes (they are Fr\'echet spaces)
$$
S^m_{\rho,\delta}(\Xi):=\left\{f\in C^\infty(\Xi)\mid\,\forall(a,\alpha),\,|f|_{(a,\alpha)}^{(m;\rho,\delta)}<\infty\right\}.
$$
We are going to work systematically under the assumption that the magnetic field $B$ has components of
class $BC^\infty(\mathcal{X})$, i.e. they are smooth and all the derivatives are bounded.
By usual oscillatory integrals techniques we prove that
for $m\in\mathbb{R}$ and $0\leq\delta\leq\rho\leq 1$ we have $S^m_{\rho,\delta}(\Xi)\subset\mathfrak{M}^B(\Xi)$.

\begin{theorem}\cite{IMP1,IMP2}

For any $m_1$ and $m_2$ in $\mathbb{R}$ and for any $0\leq\delta\leq\rho\leq 1$ we have
$$
S^{m_1}_{\rho,\delta}(\Xi)\,\sharp^B\,S^{m_2}_{\rho,\delta}(\Xi)\,\subset\,S^{m_1+m_2}_{\rho,\delta}(\Xi).
$$
\end{theorem}
Choosing any vector potential $A$ for $B$, we define the associated class of magnetic pseudodifferential operators on
 $\mathcal{H}:=L^2(\mathcal{X})$:
$$
\boldsymbol{\Psi}^m_{\rho,\delta}(A):=\mathfrak{Op}^A[S^m_{\rho,\delta}(\Xi)].
$$
Then we have
$$
\boldsymbol{\Psi}^{m_1}_{\rho,\delta}(A)\cdot\boldsymbol{\Psi}^{m_2}_{\rho,\delta}(A)\subset\boldsymbol{\Psi}^{m_1+m_2}_{\rho,\delta}(A).
$$
If $\delta=0$  we also have an asymptotic development of the composed symbol.
For any multi-index $\alpha \in \N^m$, we use the notation $\alpha!=\alpha_1! \dots \alpha_m!$.
For shortness we also set $\a:=(a,\alpha)$ and $\b:=(b,\beta)$ with $\a,\b \in \N^{2n}$.

We define $\omega_B (x,y,z):=\exp [-i\Gamma_B (x,y,z)]$, where $\Gamma_B (x,y,z)$ is the flux of $B$ through the triangle with corners
$$x-y-z,\ \ \ x+y-z,\ \ \ x-y+z.$$

\begin{theorem}\cite{LMR}

Assume that the each component $B_{jk}$ belongs to $BC^\infty(\X)$
and let $m_1,m_2 \in \R$ and $\rho \in (0,1]$. Then for any $f\in
S^{m_1}_{\rho,0}(\Xi)$, $g\in S^{m_2}_{\rho,0}(\Xi)$ and $N \in\N^*$ one has
\begin{equation*}
f\,\sharp^B g =\sum_{l=0}^{N-1}h_l + R_N,
\end{equation*}
with
\begin{equation*}
h_l = \underset{|\alpha|+|\beta|=l}{\underset{a\le\beta,
b\le\alpha}{\underset{a,b,\alpha,\beta\in\mathbb N^n}
{\sum}}}h_{\a,\b}\in S^{m_1+m_2-\rho l}_{\rho,0}(\Xi)
\end{equation*}
and
\begin{equation*}
h_{\a,\b} (x,\xi)=C_{\a\b}\big[(\partial^{\beta-a}_y\partial^{
\alpha-b}_z\omega_B)(x,0,0)\big]\,\!\big[(\partial^a_x\partial^\alpha_\xi
f)(x,\xi)\big]\,\!\big[(\partial^b_x\partial^\beta_\xi
g)(x,\xi)\big],
\end{equation*}
 the constants being given by
\begin{equation*}
C_{\a\b}=\left(\frac{i}{2}\right)^l\frac{(-1)^{|a|+|b|+
|\beta|}}{a!b!(\alpha-b)!(\beta -a)!}.
\end{equation*}
The remainder term $R_N$ belongs to $S^{m_1+m_2-\rho N}_{\rho,0}(\Xi)$.
\end{theorem}

We list the first three terms in the development:

$$h_0 =\,f g, $$

$$h_1=\frac{i}{2}\{ f,g\}=\frac{i}{2}\sum^n_{j=1}(\partial_{x_j}
f\,\partial_{\xi_j} g-\partial_{\xi_j}f\,\partial_{x_j}g),$$

$$h_2 (x,\xi)=\left(\frac{i}{2}\right)^2\underset{j,k}{\sum}
\frac{1}{\varepsilon_{jk}}\left[(\partial_{x_j}\partial_{x_k}
f)(x,\xi)(\partial_{\xi_j}\partial_{\xi_k}g)(x,\xi)+(\partial_{\xi_j}\partial_{\xi_k}f)(x,\xi
)(\partial_{x_j}\partial_{x_k}g)(x,\xi)\right]-$$

$$-\left(\frac{i}{2}\right)^2\underset{j,k}{\sum}(\partial_{x_j}\partial_{\xi_k}f)(x,\xi)(\partial_{x_k
}\partial_{\xi_j}g)(x,\xi)-\left(\frac{i}{2}\right)
\underset{j,k}{\sum}B_{jk}(x)(\partial_{\xi_k}f)(x,\xi)(\partial_{\xi_j}g)(x,\xi),$$
where $\varepsilon_{jk}=2$ if $j\neq k$ and $\varepsilon_{jj}=1$.

In \cite{IMP1} one also covers the case $0\le\delta<\rho$, but with a less explicit inductive constructions of the terms.
Developments of the magnetic product with respect to parameters can be found in \cite{Le}.

\subsection{$L^2$-continuity}

The following result can be regarded as an extension of the Calderon-Vaillancourt Theorem to the twisted Weyl calculus.

\begin{theorem}\cite{IMP1}

Assume that the magnetic field $B$ has components of class $BC^\infty(\mathcal{X})$.
In any Schr\"{o}dinger representation of the form $\mathfrak{Op}^A$,
the operator corresponding to $f\in S^0_{\rho,\rho}(\Xi)$, with $0\leq\rho<1$, defines a bounded operator
in $\mathcal H=L^2(\X)$. There exist two constants $c(n)\in\mathbb{R}_+$ and $p(n)\in\mathbb{N}$, depending only on the dimension
$n$ of the space $\mathcal{X}$, such that
$$
\|\mathfrak{Op}^A(f)\|_{\mathbb{B}(\mathcal{H})}\leq
c(n)\max\{\|\partial^a_x\partial^{\alpha}_{\xi}f\|_{\infty}\mid
|a|\leq p(n), \ |\alpha|\leq p(n)\}.
$$
\end{theorem}
Using previous notations, we can rephrase saying that $S^0_{\rho,\rho}(\Xi)\subset\mathfrak A^B(\Xi)$.

\subsection{Magnetic Sobolev spaces}

Again under the hypothesis that the magnetic field $B$ has components of class $BC^\infty(\mathcal{X})$,
we shall define the scale of Sobolev spaces starting from a special set of symbols; for any $m>0$ we define
$$
p_m(x,\xi):=<\xi>^m\equiv(1+|\xi|^2)^{m/2},
$$
so that $p_m\in S^m_{1,0}(\Xi)\subset\mathfrak{M}^B(\Xi)$. For any potential vector $A$ we set
$$
\mathfrak{p}^A_m:=\mathfrak{Op}^A(p_m).
$$
Let $A$ be a vector potential for $B$. For any $m>0$ we define the linear space
$$
\mathcal{H}_A^m(\mathcal{X}):=\left\{u\in L^2(\mathcal{X})\mid\mathfrak{p}^A_m u\in L^2(\mathcal{X})\right\}
$$
and call it {\it the magnetic Sobolev space of order $m$ associated to $A$}.

The space $\mathcal{H}_A^m(\mathcal{X})$ is a Hilbert space for the scalar product
$$
<u,v>_{(m,A)}:=<\mathfrak{p}^A_m u,\mathfrak{p}^A_m v>+<u,v>.
$$

\begin{definition}
Suppose chosen a vector potential $A$. For any $m>0$ we define the space $\mathcal{H}^{-m}_A(\mathcal{X})$ as the dual space
of $\mathcal{H}^m_A(\mathcal{X})$ with the dual norm
$$
\|\phi\|_{(-m,A)}:=\underset{u\in\mathcal{H}^m_A(\mathcal{X})\setminus\{0\}}{\sup}\frac{|<\phi,u>|}{\|u\|_{(m,A)}},
$$
that induces a scalar product. We also set $\mathcal{H}^0_A(\mathcal{X}):=L^2(\mathcal{X})$.
\end{definition}

Various properties of the scale of spaces $\{\H^m_A(\X)\mid m\in\mathbb R\}$ are studied in \cite{IMP1}. Among others,
they serve as domains for elliptic self-adjoint magnetic pseudodifferential operators.

For $m>0$ a symbol $f\in S^m_{\rho,\delta}(\Xi)$ is said to be {\it elliptic}
if there exist two positive constants $R$ and $C$ such that for $|\xi|\geq R$ one has
$$
|f(x,\xi)|\geq C<\xi>^m.
$$

\begin{theorem}\label{fifif}\cite{IMP1}

Let $B$ belong to $BC^\infty(\mathcal{X}),\, m\geq 0$, with either $0\leq\delta<\rho\leq1$ or $\delta=\rho\in[0,1)$.
 Let $f\in S^m_{\rho,\delta}(\Xi)$, real and elliptic if $m>0$.
Then for any vector potential $A$ defining $B$, the operator
$$
\mathfrak{Op}^A(f):\mathcal{H}^m_A(\mathcal{X})\rightarrow L^2(\mathcal{X})
$$
is self-adjoint in $L^2(\X)$.
\end{theorem}
If $f\geq0$ then $\mathfrak{Op}^A(f)$ is lower semibounded, by an extension of the Garding inequality to the magnetic case.

\section{Commutator criteria and applications}

\subsection{The Beals type criterion}

Let us very briefly recall \cite{Be,He} the Beals' criterion in the usual pseudodifferential calculus,
that may be obtained from our Theorem 3.1 by taking $B=0$ and $A=0$.
We introduce the following notations:
$$
\mathfrak{ad}_{Q_j}T:=Q_jT-TQ_j,\quad\mathfrak{ad}_{D_j}T:=D_jT-TD_j,\quad\forall T\in\mathbb{B}[L^2(\mathcal{X})]
$$
as sesquilinear forms on the domain of $Q_j$, resp. $D_j$.
Then $T$ has the form $T=\mathfrak{Op}(f)$, with $f\in S^0_{0,0}(\Xi)$,
if and only if for any family $\{a_1,\ldots,a_n,\alpha_1,\ldots,\alpha_n\}\in\mathbb{N}^{2n}$
the sesquilinear form
$\mathfrak{ad}^{a_1}_{Q_1}\ldots\mathfrak{ad}^{a_n}_{Q_n}\mathfrak{ad}^{\alpha_1}_{D_1}\ldots\mathfrak{ad}^
{\alpha_n}_{D_n}[T]$
defined on an obvious dense domain is continuous with respect to the $L^2(\mathcal{X})$-norm.

Given a magnetic field $B$, our purpose is to formulate a similar criterion for a bounded operator $T$ to be
in $\boldsymbol{\Psi}^{0}_{0,0}(A)$.
It is rather natural to consider the following strategy:
\begin{itemize}
\item Replace the operators $\{D_j\}_{1\leq j\leq n}$
with the magnetic momenta $\{\Pi^A_j\}_{1\leq j\leq n}$.

\item Formulate the criterion in a gauge invariant way by using the symbolic calculus developed above.
\end{itemize}
\begin{theorem}\label{ciuciu}\cite{IMP2}

 Assume that the magnetic field $B$ has components of class $BC^\infty(\mathcal{X})$.
With respect to a vector potential $A$ defining $B$, an operator $T\in\mathbb{B}[L^2(\mathcal{X})]$
has the form $T=\mathfrak{Op}^A(f)$, with $f\in S^0_{0,0}(\Xi)$,
if and only if for any family $\{a_1,\ldots,a_n,\alpha_1,\ldots,\alpha_n\}\in\mathbb{N}^{2n}$
the sesquilinear form
$\mathfrak{ad}^{a_1}_{Q_1}\ldots\mathfrak{ad}^{a_n}_{Q_n}\mathfrak{ad}^{\alpha_1}_{\Pi^A_1}\ldots\mathfrak{ad}^
{\alpha_n}_{\Pi^A_n}[T]$
is continuous with respect to the $L^2(\mathcal{X})$-norm.
\end{theorem}

In fact the above theorem is the "represented version" of a result concerning the intrinsic algebra $\mathfrak{A}^B(\Xi)$, that we shall
now present.
 In order to define the {\it 'linear monomials'} on $\Xi$ we use the canonical symplectic form $\sigma$ on $\Xi$ and consider
 for any $X\in\Xi$ the function:
$
\mathfrak{l}_X:\Xi\ni Y\mapsto\sigma(X,Y)\in\mathbb{R}.
$
 Then we can introduce {\it the algebraic Weyl system}: $\mathfrak{e}_X:=\exp\{-i\mathfrak{l}_X\}$, indexed by $X\in\Xi$.
 We define the following  twisted action of the phase space $\Xi$ by automorphisms ({\it magnetic translations}) of $\mathfrak{A}^B(\Xi)$:
$$
\Xi\ni X\mapsto\mathfrak T^B_X\in\mathbb{A}{\rm ut}[\mathfrak{A}^B(\Xi)],\quad\mathfrak T^B_X[f]:=\mathfrak{e}_{-X}\sharp^B f\sharp^
B\mathfrak{e}_X.
$$
 Some computations give $\left.-i\partial_t\mathfrak T^B_{tX}[f]\right|_{t=0}=\mathfrak{ad}^B_X[f]$,
where $\mathfrak{ad}^B_X[f]:=\mathfrak{l}_X\sharp^B f-f\sharp^B\mathfrak{l}_X$ is the magnetic derivative of $f$ in the direction $X$.

The space of $\mathfrak T^B$-regular vectors at the origin is
$$
\mathfrak{V}^{B,\infty}:=\left\{f\in\mathfrak{A}^B(\Xi)\mid \mathfrak{ad}^B_{X_1}\ldots\mathfrak{ad}^B_{X_N}[f]\in\mathfrak{A}^
B(\Xi)\right\},
$$
where $N\in\mathbb{N}$ and $\{X_1,\ldots,X_N\}\subset\Xi$ are arbitrary. The family of seminorms
$$
|f|_{X_1,\ldots, X_N}:=\|\mathfrak{ad}^B_{X_1}\ldots\mathfrak{ad}^B_{X_N}[f]\|_B
$$
indexed by all the families $\{X_1,\ldots,X_N\}\subset\Xi$, $N\in\mathbb{N}$ define on $\mathfrak{V}^{B,\infty}$ a
Fr\'echet space structure.

We also recall the usual action of $\Xi$ through translations  on the C$^*$-algebra $BC_{\rm u}(\Xi)$ (endowed with the usual norm
 $\|.\|_\infty$):
$$
\Xi\ni X\mapsto\mathfrak{T}_X\in\mathbb{A}{\rm ut}[BC_{\rm u}(\Xi)],\quad(\mathfrak{T}_X[f])(Y):=f(Y+X).
$$
The space of associated $\mathcal{T}$-regular vectors is $BC^\infty(\Xi)$ with the family of seminorms
$$
|f|_{(N)}:=\underset{|a|+|\alpha|\leq N}{\max}\|\partial^a_x\partial^\alpha_\xi f\|_\infty,
$$
indexed by $N\in\mathbb{N}$, that also induce a Fr\'echet space structure on $BC^\infty(\Xi)$.

\begin{theorem}\cite{IMP2}

 If the magnetic field $B$ has components of class $BC^\infty(\mathcal{X})$, then the two Fr\'echet spaces $\mathfrak{V}^{B,\infty}$ and
 $BC^\infty(\Xi)$ coincide (as subspaces of $\mathcal{S}^\prime(\Xi)$).
\end{theorem}

Theorem \ref{ciuciu} is a straightforward consequence of the above result
(just remark that $S^0_{0,0}(\Xi)=BC^\infty(\Xi)$).

We pass now to the case $m\ne 0$. For $(m,a)\in\mathbb{R}_+\times\mathbb{R}_+$ let $p_{m,a}(X):=a+p_m(X)=a+<\xi>^m$.
One shows that, for $a$ large enough, $p_{m,a}$ is invertible with respect to the magnetic Weyl composition law $\sharp^B$.
We denote by $p^{(-1)_B}_{m,a}\in\mathfrak A^B(\Xi)$ this inverse and set
$$
s_0:=1,
$$
$$
s_m:=p_{m,a} \quad{\rm for}\ m>0,
$$
$$
s_m:=p_{|m|,a}^{(-1)_B} \quad{\rm for}\ m<0.
$$

\begin{theorem}\label{main-m_rho}\cite{IMP2}

A distribution $f\in\mathcal{S}^\prime(\Xi)$ is a symbol of type $S^m_{\rho,0}(\Xi)$ (with $0\leq\rho\leq1$)
if and only if for any $p,q\in\mathbb{N}$ and for any $u_1,\ldots,u_p\in\mathcal{X}$ and any
$\mu_1,\ldots,\mu_q\in\mathcal{X}^*$, the following is true:
\begin{equation}\label{cond-m_rho-0}
s_{-m+q\rho}\,\sharp^B\left(\ad^B_{u_1}\cdot\ldots\cdot\ad^B_{u_p}
\ad^B_{\mu_1}\cdot\ldots\cdot\ad^B_{\mu_q}[f]\right)\in\mathfrak{A}^B(\Xi).
\end{equation}
\end{theorem}

\subsection{The Bony type criterion}

For certain purposes it is preferable to reformulate our main theorem by replacing the commutators with the linear
functions $\mathfrak{l}_X$ by more general symbols. In the absence of a magnetic field,
but for very general symbol classes defined by metrics and weights, this has been done in \cite{Bo1,BC}.

\begin{definition}
 Let $\rho\in[0,1]$; we define
$$
S^+_\rho(\Xi):=\left\{\varphi\in C^\infty(\Xi)\ \mid\ \left|\left(\partial^a_x\partial^\alpha_\xi\varphi\right)(X)\right|\leq
C_{a\alpha}<\xi>^{\rho(1-|\alpha|)},\ \text{for}\ |a|+|\alpha|\geq1\right\}.
$$
\end{definition}

For any $\varphi\in S^+_\rho(\Xi)\subset\mathfrak{M}^B(\Xi)$ we introduce the magnetic derivation
\begin{equation}
\mathfrak{ad}^B_\varphi[f]:=\varphi\,\sharp^B f-f\,\sharp^B\varphi,\qquad\forall
f\in\mathfrak{M}^B(\Xi).
\end{equation}

\begin{theorem}\label{Bony}\cite{IMP2}

$f\in\mathcal{S}^\prime(\Xi)$ belongs to $S^m_{\rho}(\Xi)$ if and only if for any
$\{\varphi_1,\ldots,\varphi_N\}\subset S^+_\rho(\Xi)$ one has
\begin{equation}\label{cond-Bony}
s_{-m}\,\sharp^B\mathfrak{ad}^B_{\varphi_1}\ldots\mathfrak{ad}^B_{\varphi_N}[f]\in\mathfrak{A}^B(\Xi).
\end{equation}
\end{theorem}

A characterization of certain classes of Fourier Integral Operators involving commutators twisted by a diffeomorphism
can be found in \cite{Bo2,Bo3}. The magnetic case (but for restricted metrics and weights) is considered in \cite{IMP2}.
Using this, one can prove that
the evolution group of a (rather restricted) class of magnetic pseudodifferential operator belongs to a natural class of
magnetic FIO associated to the classical flow of the symbol.

\subsection{Inversion}

By using the magnetic Bony criterion and the behavior of inversion under products of derivations of the form $\mathfrak{ad}^B_{\varphi}$,
it is not difficult to prove

\begin{proposition}\label{th}\cite{IMP2}

If $f\in S^0_{\rho,0}(\Xi)$ is invertible in the $C^*$-algebra $\mathfrak{A}^B(\Xi)$,
then the inverse $f^{(-1)_B}$ also belongs to $S^0_{\rho,0}(\Xi)$.
\end{proposition}

We recall (cf. \cite{LaMN} and references therein) that a $\Psi^*$-{\it algebra}
is a Fr\'echet $^*$-algebra continuously embedded in a $C^*$-algebra, which is
spectrally invariant (i.e. stable under inversion). Our Proposition \ref{th}
says that $S^0_{\rho,0}(\Xi)$ is a $\Psi^*$-algebra in the $C^*$-algebra $\mathfrak A^B(\Xi)$.

By same simple abstract nonsense one extends the result to unbounded symbols:

\begin{proposition}\label{symb-inv}
Let $m>0$ and $\rho\in[0,1]$.
If $g\in S^m_\rho(\Xi)$ is invertible in $\mathfrak M^B(\Xi)$, with $\mathfrak s_{m}\sharp^B g^{(-1)_B}\in\mathfrak A^B(\Xi)$,
then $g^{(-1)_B}\in S^{-m}_\rho(\Xi)$.
\end{proposition}
 We can apply Proposition \ref{symb-inv} to elliptic symbols of strictly
positive order by using Theorem \ref{fifif}. The spectrum $\sigma[f]$ of the operator $\mathfrak{Op}^A$ does not
depend on the choice of $A$ (by gauge covariance). Thus, for any $z\notin\sigma[f]$,
the operator $\mathfrak{Op}^A(f)-z1=\mathfrak{Op}^A(f-z)$ is invertible with bounded inverse. This
means that the inverse $(f-z1)^{(-1)_B}$ exists in $\mathfrak{M}^B(\Xi)$ and belongs to
$\mathfrak{A}^B(\Xi)$. It is easy to show that
$p_m\sharp^B(f-z)^{(-1)_B}\in\mathfrak{A}^B(\Xi)$. This allows us to conclude:

\begin{proposition}\label{ell-inv}\cite{IMP2}

Given a real elliptic symbol $f\in S^m_{\rho,0}(\Xi)$, for any $z\notin\sigma[f]$ the inverse $(f-z)^{(-1)_B}$ exists and it is a symbol
of class $S^{-m}_{\rho,0}(\Xi)$.
\end{proposition}

\subsection{Functional calculus}

Propositions \ref{th} and \ref{ell-inv} imply results concerning the functional
calculus of elliptic magnetic self-adjoint operators. The formula
$$
\Phi\left(\mathfrak {Op}^A[f]\right)=:\mathfrak {Op}^A\left[\Phi^B(f)\right]
$$
gives an intrinsic meaning to the functional calculus for Borel functions $\Phi$.

First, $\Psi^*$-algebras are stable under the holomorphic functional
calculus, so we can state as a consequence of Proposition \ref{th}:

\begin{proposition}\label{zbauer}\cite{IMP2}

If $f\in S^0_{\rho,0}(\Xi)$ and $\Phi$ is a function holomorphic on
some neighborhood of the spectrum of $f$, then $\Phi^B(f)\in S^0_{\rho,0}(\Xi)$.
\end{proposition}

If $\Phi\in C_0^\infty(\R)$, then $\Phi^B(f)$ can be written using the
Helffer-Sj\"ostrand formula
\begin{equation}\label{hesj}
\Phi^B(f)=\frac{1}{\pi}\int_\mathbb C dz\,\partial_{\overline z}\widetilde{\Phi}(z)(f-z)^{(-1)_B},
\end{equation}
$\widetilde{\Phi}$ being a quasi-analytic extension of $\Phi$ (cf. \cite{HS}). This allows applying Proposition \ref{ell-inv} and one gets
rather straightforwardly:

\begin{theorem}\label{func-calc}\cite{IMP2}

If $\Phi\in C^\infty_0(\mathbb{R}), f\in S^m_{\rho,0} (\Xi)$, elliptic if $m>0$, then $\Phi^B(f)\in S^{-m}_{\rho,0}(\Xi)$.
\end{theorem}

Choosing a vector potential $A$ for the magnetic field $B$, one proves the following result about the fractional powers of the
operator $\mathfrak{Op}^A(f)$:

\medskip
\begin{theorem}\label{codasa}\cite{IMP2}

 Given a lower bounded $f\in S^m_{\rho,0}(\Xi)$ with $m\geq0$, elliptic if $m>0$, let $\mathfrak{Op}^A[f]$ be the
 associated self-adjoint, semi-bounded operator on $\mathcal{H}$ given by Theorem \ref{fifif}.
 Let $t_0\in\mathbb{R_+}$ such that for $f_0:=f+t_01$ the operator $\mathfrak{Op}^A[f_0]$ is strictly positive. Then
 for any $s\in\mathbb{R}$ the power $s$ of $\mathfrak{Op}^A[f_0]$ is a magnetic pseudodifferential operator
 with symbol $f_0^{[s]_B}\in S^{sm}_\rho(\Xi)$, i.e. $\left(\mathfrak{Op}^A[f_0]\right)^s=
 \mathfrak{Op}^A\big[f_0^{[s]_B}\big]$.
\end{theorem}

\section{$C^*$-algebras of magnetic symbols and operators}

\subsection{Twisted crossed products}

Starting from a magnetic twisted $C^*$-dynamical system, we shall now reconstruct $C^*$-algebras of magnetic symbols \cite{MPR1}.
These are particular instances of twisted $C^*$-algebras introduced in \cite{PR1} and \cite{PR2}.

We call {\it admissible algebra} a unital $C^*$-subalgebra $\A$ of $BC_{{\rm u}}(\X)$ which is invariant under translations.
Consequently, denoting by $S_\A$ the Gelfand spectrum of $\A$, the map $\theta: \X \times \X \to \X,\ \theta(x,y):=x+y$
extends to a continuous map $\theta: \SA \times \X \to \SA$. We also use the notations $\theta(\kappa,y)=\theta_y(\kappa)=\theta^\kappa(y)$
for $(\kappa,y)\in\SA\times\X$ and get a topological dynamical system $(\SA,\theta,\X)$ with compact space $\SA$.
By duality, we also call $\theta$ the action of $\X$ in $\A$.

Now assume that the components $B_{jk}$ of the magnetic field belong to $\A$. We define for each $x,y,z \in \X$
\begin{equation*}
\omega^B(z;x,y) := e^{- i \Gamma^B(<z,z+x,z+x+y>)},
\end{equation*}
$\Gamma^B(<a,b,c>)$ denoting the flux of $B$ through the triangle $<a,b,c>$.
For fixed $x$ and $y$, the function $\omega^B(\cdot;x,y)\equiv \omega^B(x,y)$ belongs to the unitary group $\U(\A)$
of $\A$. Moreover, the mapping $\X \times \X \ni (x,y) \mapsto \omega^B(x,y) \in \U(\A)$
is a continuous normalized 2-cocycle on $\X$, i.e. for all $x,y,z \in \X$ the following
hold (as a consequence of Stokes' Theorem):
\begin{equation}\label{eq2cocycle}
\omega^B(x+y,z) \, \omega^B(x,y) = \theta_x[\omega^B(y,z)] \, \omega^B(x,y+z),
\end{equation}
\begin{equation*}
\omega^B(x,0)=\omega^B(0,x)=1.
\end{equation*}

Let $L^1(\X; \A)$ be the set of Bochner integrable functions on $\X$ with
values in $\A$, with the $L^1$-norm $\|F \|_1 :=\int_{\X} \de
x\;\!\|F(x)\|_\A$. For any $F,G \in L^1(\X;\A)$ and $x \in \X$, we define the product
\begin{equation*}
(F \diamond^B G)(x):=\int_{\X}\de y\;\theta_{\frac{y-x}{2}}\!\left[F(y)
\right]\;\!\theta_{\frac{y}{2}}\!\left[G(x-y)\right]\;\!\theta_{-\frac{x}{2}}\!\left[\oB(y,x-y)\right]
\end{equation*}
and the involution
\begin{equation*}
F^{\diamond^B}(x):=\overline{F(-x)}.
\end{equation*}

\begin{definition}\label{primel}
The enveloping $C^*$-algebra of $L^1(\X;\A)$ is called {\it the twisted crossed
product} and is denoted by $\A\!\rtimes^{\oB}_\theta\!\!\X$, or simply by $\CBA$.
\end{definition}

\begin{theorem}\cite{MPR1}
The partial Fourier transform $\mathcal F\otimes 1$ defines by extension a $C^*$-isomorphism $\mathfrak F:\CBA\rightarrow\BBA$,
where $\BBA$ is a $C^*$-subalgebra of $\mathfrak A^B(\Xi)$.
\end{theorem}

There is a one-to-one correspondence between covariant representations of a twisted $C^*$-dynamical system and non-degenerate
representations of the twisted crossed product. We denote by $\U(\H)$ the group of unitary operators in the Hilbert space $\H$.

\begin{definition}\label{RCT}
Given a magnetic $C^*$-dynamical system $(\A,\theta,\oB,\X)$, we call {\it
covariant representation} $(\H,r,T)$ a Hilbert space $\H$ together with two
maps $r:\A\rightarrow \B(\H)$ and $T:\X\rightarrow \U(\H)$ satisfying:
\begin{itemize}
\item $r$ is a non-degenerate representation,
\item $T$ is strongly continuous
and $$T(x)\;\!T(y)=r[\oB(x,y)]\;\!T(x+y), \quad \forall x,y\in \X,$$
\item $T(x)\;\!r(a)\;\!T(x)^*=r[\theta_x(a)], \quad \forall x\in \X,\;a\in\A$.
\end{itemize}
\end{definition}

\begin{lemma} If $(\H,r,T)$ is a covariant representation of
$(\A,\theta,\oB,\X)$, then $\Rep^T_r$ defined on $L^1(\X;\A)$ by
\begin{equation*}
\Rep_r^T(F):=\int_\X \de x\,r\left[\theta_{\frac{x}{2}}
\big(F(x)\big)\right]T(x)
\end{equation*}
extends to a representation of $\CBA=\A\rtimes^{\oB}_\theta\!\!\X$.
\end{lemma}

By composing with the partial Fourier transformation, one gets the most general
representations of the pseudodifferential $C^*$-algebra $\BBA$, denoted by
\begin{equation}\label{gigi}
\mathfrak {Op}^T_r:\BBA\rightarrow\B(\H),\ \ \ \ \ \mathfrak {Op}^T_r(f):=\Rep^T_r[\FF^{-1}(f)].
\end{equation}
To get the representation $\mathfrak {Op}^A$  (with $dA=B$) by this procedure one stars with the covariant representation
$L^2(\X, r, T^A)$, where
$$[r(\varphi)u](x):=\varphi(x)u(x),\ \ \forall x\in\X,\ \ \forall u\in L^2(\X),\ \ \forall \varphi\in\mathcal A$$
and {\it the magnetic translations} $T^A(y)$ are given by
$$[T^A(y)u](x):=\exp \left[-i\int^{x+y}_x A\right]u(x+y),\ \ \forall x,y\in\X,\ \ \forall u\in L^2(\X).$$

\subsection{Symbol classes with coefficients in a $C^*$-algebras}

We now introduce the anisotropic version of the H\"ormander classes of
symbols and show their relationship with twisted crossed products. For any $f:\Xi\rightarrow \C$ and $(x,\xi)\in\Xi$,
we will often write $f(\xi)$ for $f(\cdot, \xi)$ and $[f(\xi)](x)$ for $f(x,\xi)$. Thus, $f$ will
be seen as a function on $\X^*$ taking values in some space of functions
defined on $\X$. As before, $\A$ will be an admissible algebra.
The next definition is adapted from \cite{Ba2}, see also \cite{Ba1,CMS,Co,Sh}.

\begin{definition}
The space of \emph{$\A$-anisotropic symbols of order $m$ and type $(\rho,\delta)$} is
\begin{equation}\label{symbol}
S^m_{\rho,\delta}(\X^*;\A^\infty) = \bigl \{ f \in
S^m_{\rho,\delta}(\Xi) \mid  (\partial^a_x\partial^\alpha_\xi f)(\xi)
\in \A , \ \forall \xi\in\X^* \hbox{ \rm and } \alpha,a\in\mathbb N^n\bigr \}.
\end{equation}
\end{definition}
In particular, for $\A={BC_u}(\X)$, it is easy to see that
 $S^m_{\rho,\delta} \bigl ( \X^*;BC_u(\X)^{\infty} \bigr)=S^m_{\rho,\delta}(\Xi)$.

One shows the following properties, cf. \cite{LMR}:
\begin{itemize}
\item $S^{m}_{\rho,\delta}(\X^*,\A^\infty)$ is a closed subspace of the Fr\'echet
 space $S^{m}_{\rho,\delta}(\Xi)$.
 \item For any $\alpha,a \in \N^n$, $\partial^a_x\partial^\alpha_\xi S^{m}_{\rho,\delta}(\X^*,\A^\infty)
\subset S^{m-\rho |\alpha|+\delta |a|}_{\rho,\delta}(\X^*,\A^\infty)$.
\item For any $m_1, m_2 \in \R$, $$S^{m_1}_{\rho,\delta}(\X^*,\A^\infty)\cdot
  S^{m_2}_{\rho,\delta}(\X^*,\A)
\subset S^{m_1+m_2}_{\rho,\delta}(\X^*,\A^\infty)$$ and
\begin{equation}\label{eqmain}
S^{m_1}_{\rho,\delta}(\X^*,\A^\infty)\;\sharp^B\; S^{m_2}_{\rho,\delta}(\X^*,\A^\infty)\;\subset\;
S^{m_1+m_2}_{\rho,\delta}(\X^*,\A^\infty).
\end{equation}
\end{itemize}

\begin{proposition}
The $C^*$-algebra $\BBA$ is generated by anyone of the following subsets:
\begin{itemize}
\item $S^{m}_{\rho,0}(\X^*,\A^\infty)$, for some $m<0,\,\rho\in[0,1]$, cf. \cite{LMR}.
\item $\{\varphi\,\sharp^B \psi\mid \varphi\in\A,\,\psi\in \S(\X^*)\}$ or by $\{\psi\,\sharp^B \varphi\mid \psi\in\A,\,\psi\in \S(\X^*)\}$,
cf. \cite{MPR1}.
\end{itemize}
\end{proposition}

\subsection{Inversion and affiliation}

As explained above, $S^0_{\rho,0}(\Xi)$  is a $\Psi^*$-algebra in $\mathfrak A^B(\Xi)$.
Then, since $S^0_{\rho,0}(\T)$ is a closed $^*$-subalgebra of
$S^0_{\rho,0}(\Xi)$, by a result in \cite{La}, it
follows that $S^0_{\rho,0}(\T)$ is also a $\Psi^*$-algebra in $\mathfrak A^B(\Xi)$. In particular,
if $f \in S^0_{\rho,0}(\T)$ has an inverse in $\mathfrak A^B(\Xi)$ with respect to $\sharp^B$, then this inverse belongs to
$S^0_{\rho,0}(\X^*,\A^\infty)$. As by-products of the theory of $\Psi^*$-algebras, one can state
\begin{proposition}
$S^0_{\rho,0}(\T)$ is a $\Psi^*$-algebra, it is stable under the holomorphic
functional calculus, the group of invertible elements is open and the map
$[S^0_{\rho,0}(\T)]^{(-1)_B}\ni f\mapsto f^{(-1)_B}\in S^0_{\rho,0}(\T)$ is continuous.
\end{proposition}
Once again by a simple use of the symbolic calculus, one gets the extension to unbounded symbols:

\begin{theorem}\label{version}\cite{LMR}

Let $m>0$, $\rho \in [0,1]$ and $f$ be a real-valued elliptic element of
$S^m_{\rho,0}(\T)$. Then for any $z \in\C\setminus\R$, the function
$f-z$ is invertible with respect to $\sharp^B$ and its inverse $(f-z)^{(-1)_B}$ belongs to $S^{-m}_{\rho,0}(\T)$.
\end {theorem}

The main application of this Theorem is to connect unbounded symbols (and operators) to the twisted crossed product algebras.
For this we borrow a concept from \cite{ABG}.

\begin{definition}\label{secundel}
\emph{An observable affiliated to a $C^*$-algebra} $\CCC$ is a morphism $\Phi:
C_0(\R) \to \CCC$.
\end{definition}

For example, if $\H$ is a Hilbert space and $\CCC$ is a $C^*$-subalgebra
of $\B(\H)$, then a self-adjoint operator $H$ in $\H$ defines an observable
$\Phi_{\!\hbox{\it \tiny H}}$ affiliated to $\CCC$ if and only if
$\Phi_{\!\hbox{\it \tiny H}}(\eta) := \eta(H)$ belongs to $\CCC$ for all $\eta
\in C_0(\R)$. A sufficient condition is that $(H-z)^{-1} \in \CCC$ for some $z
\in \C$ with $\im z \neq 0$. Thus an observable affiliated to a $C^*$-algebra
is the abstract version of the functional calculus of a self-adjoint operator.

The next Theorem is a rather simple corollary of our previous results and will be used subsequently in the  spectral
analysis of magnetic pseudodifferential operators.

\begin{theorem}\label{afileisn}\cite{LMR}

For $m>0$ and $\rho \in [0,1]$, any real-valued elliptic element $f\in
S^m_{\rho,0}(\T)$ defines an observable affiliated to the $C^*$-algebra $\BBA$.
\end{theorem}

\section{Applications to spectral analysis}

\subsection{The limiting absorption principle}

This subsection follows Section 7 in \cite{IMP1} and is devoted to the spectral analysis and a
limiting absorption principle of operators of the form
$\mathfrak{Op}^A(f)$ for an elliptic symbols $f\in S^m_{1,0}(\Xi)$.

\begin{hypothesis}\label{I}
There exists $\epsilon>0$ such that for any
$\alpha\in\mathbb{N}^n$ there exists $C_\alpha>0$ with $$|(\partial^\alpha B_{jk})(x)|\leq C_\alpha<x>^{-1-\epsilon},$$
for any $x\in\mathcal{X}$ and $j,k\in \{1,\ldots,n\}$.
\end{hypothesis}

We shall denote by $\mathfrak g$ the metric $\mathfrak g_{_X}:=|dx|^2+<\xi>^{-2}|d\xi|^2$ and by $M_{m,\delta}$
the weight function $M_{m,\delta}(X):=<x>^{-\delta}<\xi>^m$, for $X=(x,\xi)\in\Xi$.

We recall that $f\in S^m_{1,0}(\Xi)$ is a real elliptic symbol and $m>0$, then the operator $H:=\mathfrak{Op}^A(f)$
is self-adjoint in $L^2(\mathcal{X})$.

\begin{hypothesis}\label{II}

\begin{itemize}
\item $f\in S^m_{1,0}(\Xi)$ is real valued and elliptic and can be written in the form $f=f_0+f_S+f_L$.
\item $f_0\in S^m_{1,0}(\Xi)$ a real elliptic symbol depending only on the variable $\xi\in\mathcal{X}^*$, and it is
positive for $|\xi|$ large.
\item The symbol $f_S$ of the short-range perturbation belongs to $S(M_{m,1+\epsilon},\mathfrak g)$.
\item The symbol $f_L$ of the long-range perturbation belongs to $S(M_{m-1,\epsilon},\mathfrak g)$.
\end{itemize}
\end{hypothesis}

For $t$ and $s$ in $\mathbb{R}$, let us denote by $\mathcal{H}^s_t$ the usual weighted Sobolev space:
$$
\mathcal{H}^s_t=\{u\in\mathcal S'(\mathcal{X})\mid\;<D>^s<Q>^tu\in L^2(\mathcal{X})\}.
$$
We also define $\mathbb C_\pm:=\{z\in\mathbb C\mid \pm\im(z)>0\}$.

\begin{theorem}\label{89}\cite{IMP1}

Assume that the magnetic field $B$ satisfies Hypothesis \ref{I}. Let $f\in S^m_{1,0}(\Xi)$, with $m>0$, satisfying Hypothesis \ref{II}.
Let $H$, $H_0$, respectively, the self-adjoint operators defined by $\mathfrak{Op}^A(f)$ and $\mathfrak{Op}^A(f_0)$ in
$\H:=L^2(\mathcal X)$. They have the following properties:
\begin{enumerate}
\item[a)] $\sigma_{\text{\sl ess}}(H)=\sigma_{\text{\sl ess}}(H_0)=\overline{f_0(\mathcal{X^*})}$.
\item[b)] The singular continuous spectrum of $H$ (if it exists) is contained in the set of critical values of $f$, defined as
${\rm Cr}(f_0):=\{f_0(\xi)\;\mid\;f^\prime_0(\xi)=0\}$.
\item[c)] The eigenvalues of $H$ outside ${\rm Cr}(p_0)$ have finite multiplicity and can accumulate only in ${\rm Cr}(f_0)$.
\item[d)] {\sl (Limiting Absorption Principle)} For any $\gamma>\frac{1}{2}$, the holomorphic function
$\mathbb{C}_\pm\ni z\mapsto (H-z)^{-1}\in\mathcal{B}
(\mathcal{H}^{-m/2}_\gamma,\,\mathcal{H}^{m/2}_{-\gamma})$ has a continuous extension to $\overline{\mathbb{C}_\pm}\setminus
\left[{\rm Cr}(f_0)\cup\sigma_{\text{\sl p}}(H)\right]$.
\end{enumerate}
\end{theorem}

The main tool to prove this result is Mourre theory in the form presented in \cite{ABG},
combined with the magnetic pseudodifferential calculus developed above. The limiting absorption theorem can be improved
by using real interpolation spaces, as in \cite{ABG}.
Weaker and less general results on the spectral analysis of operators of the form $\mathfrak{Op}^A(f)$ were obtained in \cite{Pa,Um}.
Even the simple Schr\"odinger case ($f_0(\xi)=|\xi|^2,\,f_S=0=f_L$) shows that $\epsilon\le 0$ is forbidden in Hypothesis \ref{I}.

\subsection{Essential spectrum of anisotropic operators}

We consider now self-adjoint operators $\mathfrak{Op}^A(f)$ in $L^2(\X)$, defined by a real and elliptic symbol $f\in S^m_{\rho,0}(\T)$
and by a magnetic field $B=dA$ with components in $\A^\infty$. The structure of the essential spectrum of such an operator can
be read in $\mathbf Q_\A$, the set of all quasi-orbits (orbit closures) of the topological dynamical
system $(\SA,\theta,\X)$.

To explain this, we follow the strategy of \cite{M,MPR2,LMR} (see also references therein).
We are going to assume that $\A$ contains $C_0(\X)$, so $\SA$
is a compact space and $\X$ can be identified with a dense open subset of $\SA$.
The complement $\FA$ of $\X$ in $\SA$ is closed and
invariant; it is the space of a compact topological dynamical system.

Any $\varphi\in\A$ extends to a continuous function $\tilde\varphi:S_\A\rightarrow\C$. Then, for any quasi-orbit $\mathcal Q$,
$\tilde\varphi$ restricts to an element $\tilde\varphi|_\mathcal Q$ of $C(\mathcal Q)$. To reinterprete $\tilde\varphi|_\mathcal Q$
as a function on $\X$, we choose an element $\kappa\in\mathcal Q$ such that $\{\theta_x[\kappa]\equiv\theta^\kappa[x]\mid x\in\X\}$
is dense in $\mathcal Q$. We define $\varphi_\mathcal Q:\X\rightarrow\C$ by
$\varphi_\mathcal Q:=\tilde\varphi|_\mathcal Q\circ\theta^\kappa$, i.e.
$$
\varphi_\mathcal Q(x):=\tilde\varphi\left(\theta_x[\kappa]\right),\ \ \ \ \ \forall x\in\X
$$
and check easily that it is a bounded, uniformly continuous function.

To sum up, we have defined a $^*$-morphism
$$
\A\ni\varphi\mapsto\varphi_\mathcal Q\in BC_{{\rm u}}(\X),
$$
whose range is a $C^*$-subalgebra $\A_\mathcal Q$ of $BC_{{\rm u}}(\X)$.

Since the components $B_{jk}$ of the magnetic field belong to $\A^\infty\subset\A$, one gets for any $\mathcal Q\in\mathbf Q_\A$
a smooth magnetic field $B_\mathcal Q$ having all the derivatives in $\A_\mathcal Q$.

The same is true for $f(\cdot,\xi)$ for any fixed $\xi\in\X^*$, and it comes out that the function
$$
f_\mathcal Q(x,\xi):=\left[f(\cdot,\xi)_\mathcal Q\right](x),\ \ \ \ \ \forall(x,\xi)\in\Xi
$$
is an elliptic element of $S^m_{\rho,0}(\X^*;\A_\mathcal Q^\infty)$ (obvious definition).

Now one only needs to choose a family $\mathbf Q$ of quasi-orbits covering $F_\A:=S_\A\setminus\X$ and for each $\mathcal Q\in\mathbf Q$
a vector potential $A_\mathcal Q$ such that $dA_\mathcal Q=B_\mathcal Q$. Using all these, one can state

\begin{theorem}\label{thmess}\cite{LMR}

Let $m>0$, $\rho \in [0,1]$ and let $\mathbf Q\subset \mathbf Q_\A$ define a covering of $\FA$ with quasi-orbits.
Then, for any magnetic field $B=dA$ of class $\A^\infty$ and for any real-valued elliptic element $f$ of $S^m_{\rho,0}(\T)$, one has
\begin{equation}\label{cici}
\sigma_{\hbox{\rm \tiny ess}}\big[\mathfrak{Op}^A(f)\big] = \overline{\bigcup_{\mathcal Q \in \mathbf Q}
\sigma[\mathfrak {Op}^{A_\mathcal Q}(f_\mathcal Q)]}.
\end{equation}
\end{theorem}

The choice of vector potentials serve only to express the result in a conventional way. The proof is completely
intrinsic, consisting only in direct manipulations of the symbols $f$ (affiliated to the $C^*$-algebra $\mathfrak B^B_\A$) and
$f_\mathcal Q$ (affiliated to the $C^*$-algebra $\mathfrak B^{B_\mathcal Q}_{\A_\mathcal Q}$). Even the statement could only
involve $(f,B)$ and $(f_\mathcal Q,B_\mathcal Q)_{\mathcal Q\in\mathbf Q}$, at the cost of introducing some abstract notions involving
spectra of observables affiliated to $C^*$-algebras. We refer to \cite{LMR} for further explanations and a full proof.

\begin{remark}\label{prak}
Combining our approach with techniques from \cite{ABG,GI1,GI3}, one could extend the result above to more singular symbols $f$.
Examples of algebras $\A$ for which the quasi-orbit structure of the Gelfand spectrum is explicit can be found in
\cite{ABG,AMP,GI1,GI2,GI3,M,MPR2,Ri} and will not be reviewed here.
Non-propagation results easily follow by adapting to the present general framework the approach
in \cite{AMP,MPR2}. Other general results for the essential spectrum of self-adjoint operators with or without magnetic fields
are included in \cite{HM,LS,LN,Ra,RRS}.
\end{remark}

\bibliographystyle{amsalpha}

\end{document}